\documentclass[conference, 10pt]{IEEEtran}
\ifCLASSINFOpdf
\else
\fi
\usepackage{graphicx,subfigure}
\usepackage{color}
\usepackage{placeins}
\usepackage{float}
\usepackage{hyperref}
\usepackage{tabularx,colortbl}

\begin{document}
\title{Self-Oscillating Capacitive Wireless Power Transfer \\with Robust Operation}
\author{
	\IEEEauthorblockN{Fu Liu$^*$, Bhakti Chowkwale, Sergei A. Tretyakov}
	\IEEEauthorblockA{Department of Electronics and Nanoengineering\\
		Aalto University, P.O. Box 15500, FI-00076 Aalto, Finland\\ 
		Email: fu.liu@aalto.fi}
}

\maketitle

\begin{abstract}
We show that a capacitive wireless power transfer device can be designed as a self-oscillating circuit using operational amplifiers. As the load and the capacitive wireless channels are part of the feedback circuit of the oscillator, the wireless power transfer can self-adjust to the optimal condition under the change of the load resistance and the transfer distance. We have theoretically analyzed and experimentally demonstrated the proposed design. The results show that the operation is robust against changes of various parameters, including the load resistance.
\end{abstract}
\IEEEpeerreviewmaketitle

\section{Introduction}
In the recent decade, wireless power transfer (WPT) became  a remarkable research topic due to its vast real-life application potential. In the conventional WPT systems, including both inductive and capacitive ones, the power generator and the power receiver are designed to have the same resonance frequency to obtain high efficiency \cite{Dai2015survey}. Unsatisfactorily, these systems are usually non-robust as the working band is narrow and dynamic tuning is required in actual implementations due to changes of the receiver parameters. In 2017, Ref.~\cite{Fan2017PT} obtained robust WPT in an inductive system by using the parity-time symmetry concept with balanced gain and loss. In the same year, Ref.~\cite{Radi2017arxiv} proposed a more general self-oscillating regime for robust operation in both capacitive and inductive WPT systems. However, a real implementation of the capacitive WPT based on a self-oscillating circuit is still lacking. In this work, we introduce one design with an operational amplifier (Op-Amp) and theoretically and experimentally demonstrate the robustness of such new WPT devices.

\section{Self-Oscillating Circuit Based \\Capacitive WPT}
\begin{figure}[t!]
	\centering
	\noindent
	\includegraphics[width=0.22\textwidth]{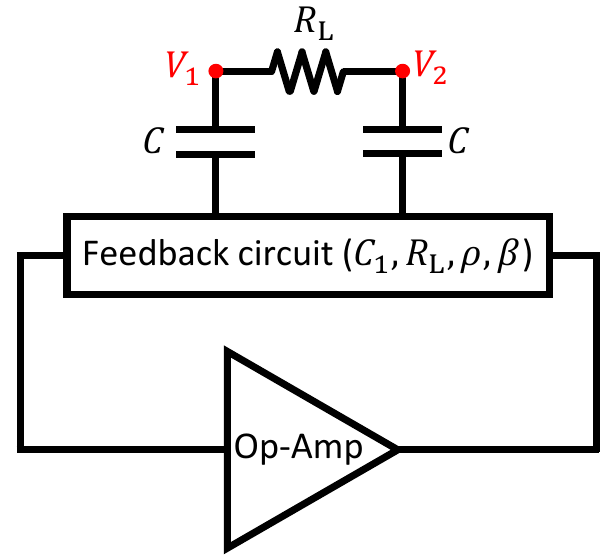}
	\caption{Schematic of the self-oscillating circuit designed with Op-Amp for capacitive WPT. $C_1$, $R_{\rm{L}}$, $\rho$, and $\beta$ are the feedback parameters.}\label{fig1}
\end{figure}

The proposed capacitive WPT design based on a self-oscillating circuit with an Op-Amp is schematically shown in Fig.~\ref{fig1}. In this design, self-oscillations are sustained by the feedback circuit which also includes the load resistor $R_{\rm{L}}$ and the two capacitors $C$. Here, the two capacitors $C$ provide  wireless channels as we can replace them by parallel metal plates in actual WPT implementations. Due to self-oscillations, the power can be wirelessly delivered into the load, even when $R_{\rm{L}}$ or $C$ are changing, as long as the self-oscillating condition is satisfied. Therefore, this WPT design is robust to the changes of the receiver properties, $R_{\rm{L}}$ and $C$.

The feedback of the self-oscillating circuit is characterized by the parameters $C_1$, $R_{\rm{L}}$, $\rho$, and $\beta$, in which the wireless channel property $C$ is embedded in $\rho$. This circuit can be analyzed theoretically and we have obtained two voltages $V_1$ and $V_2$ as functions of time $t$. Based on these analytical results, the oscillating period is obtained as
\begin{equation}
	T=2C_1 R_{\rm{L}} \frac{\rho}{2+\rho} \ln{\frac{\rho+\beta(2+\rho)}{\rho-\beta(2+\rho)}},
	\label{eqT}
\end{equation}
while the self-oscillating condition reads
\begin{equation}
	\beta<\frac{\rho}{2+\rho}.
	\label{eqcond}
\end{equation}

On the other hand, the averaged power $P$ delivered to the load can be calculated by the formula $P=\frac{1}{T R_{\rm{L}}} \int_0^T (V_2-V_1)^2 dt$, which gives the power ratio 
\begin{equation}
	P_{\rm{r}}=\frac{P}{P_0}
	=\frac{2\beta(2+\rho)}{\rho}\bigg/\ln{\frac{\rho+\beta(2+\rho)}{\rho-\beta(2+\rho)}},
	\label{eqPr}
\end{equation}
where $P_0=V_{\rm{CC}}^2/R_{\rm{L}}$ is the power consumption on the load if the DC source $V_{\rm{CC}}$ of the Op-Amp is directly applied on the load. We note that this power ratio is only a function of $\beta$ and $\rho$. From these equations, it is clear that the proposed WPT can operate in a wide range of feedback parameters.

\section{Experimental Results}
We have experimentally implemented the proposed WPT design with an Op-Amp model TL072IP from Texas Instruments. As the main purpose is to demonstrate the robustness of the proposed WPT design, we use lumped capacitors $C$ instead of real metal plates in the implementation. In the following, we will present the robustness of our design by showing $T$ and $P_{\rm{r}}$ while varying the feedback parameters. 

\subsection{Varying Feedback Parameter $\beta$}
\begin{figure}[t!]
	\centering
	\subfigure[]
	{\includegraphics[width=0.24\textwidth]{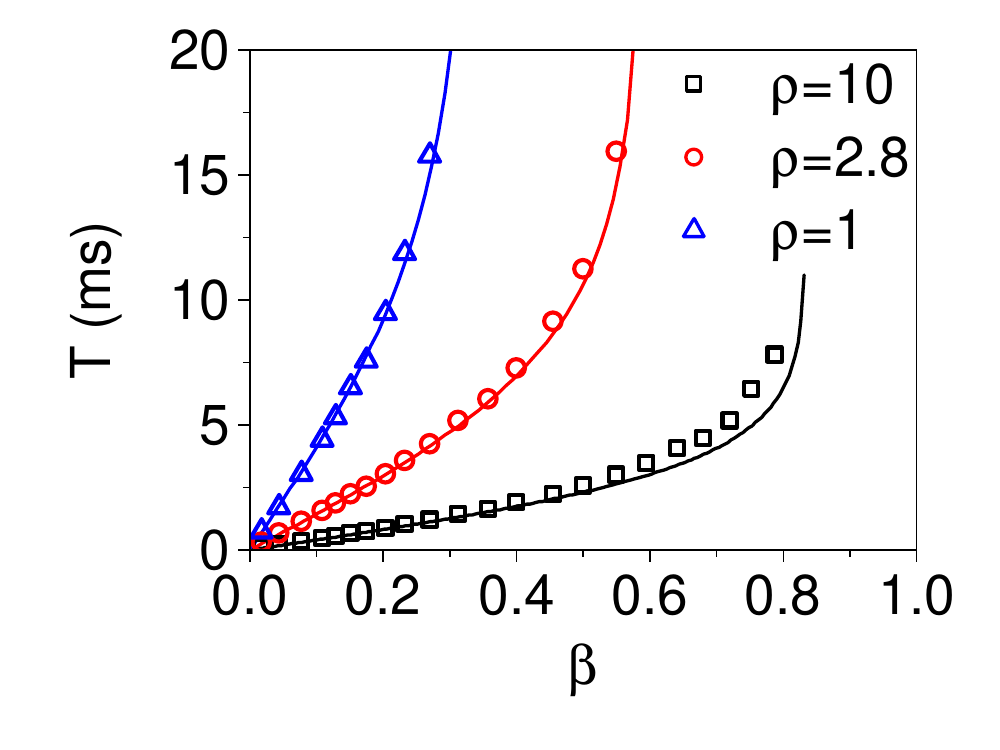}\label{fig2a}}
	\centering
	\subfigure[]
	{\includegraphics[width=0.24\textwidth]{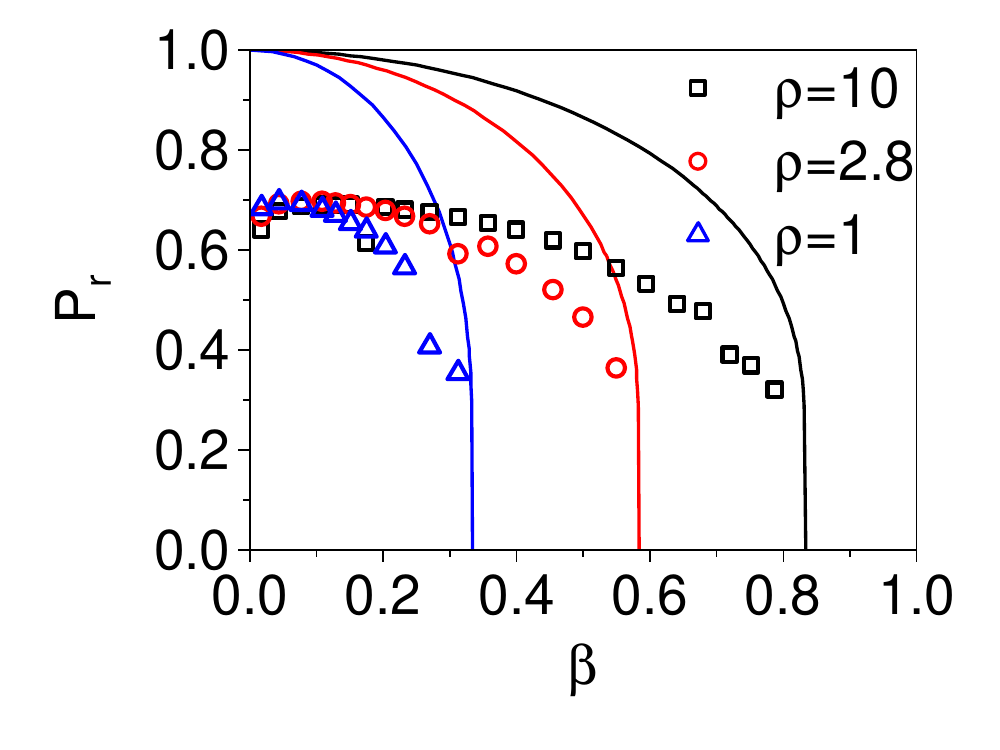}\label{fig2b}}
	\caption{(a) Oscillation period $T$ and (b) power ratio $P_{\rm{r}}$ on the load when varying $\beta$ for different $\rho$ and $C_1$ configurations.}
	\label{fig2}
\end{figure}

First of all, we fix the load resistance $R_{\rm{L}}=10~\rm{k\Omega}$ and study how the oscillating period $T$ and the power ratio $P_{\rm{r}}$ behave when we change the feedback parameter $\beta$. As examples, we have chosen three sets of $\rho$ and $C_1$ configurations: $\rho=10$ with $C_1=100$~nF, $\rho=2.8$ with $C_1=356$~nF, and $\rho=1$ with $C_1=1$~uF. The corresponding experimental results are shown as open symbols in Fig.~\ref{fig2}, while the curves are from analytical formula Eq.~\ref{eqT} and \ref{eqPr}. From Fig.~\ref{fig2a}, it is obvious that the experimentally measured period $T$ follows the analytic formula very well. However, in Fig.~\ref{fig2b}, the measured power ratio $P_{\rm{r}}$ is smaller than the expected values, meaning that less power is delivered to the load.

\begin{figure}[h]
	\centering
	\noindent
	\includegraphics[width=0.24\textwidth]{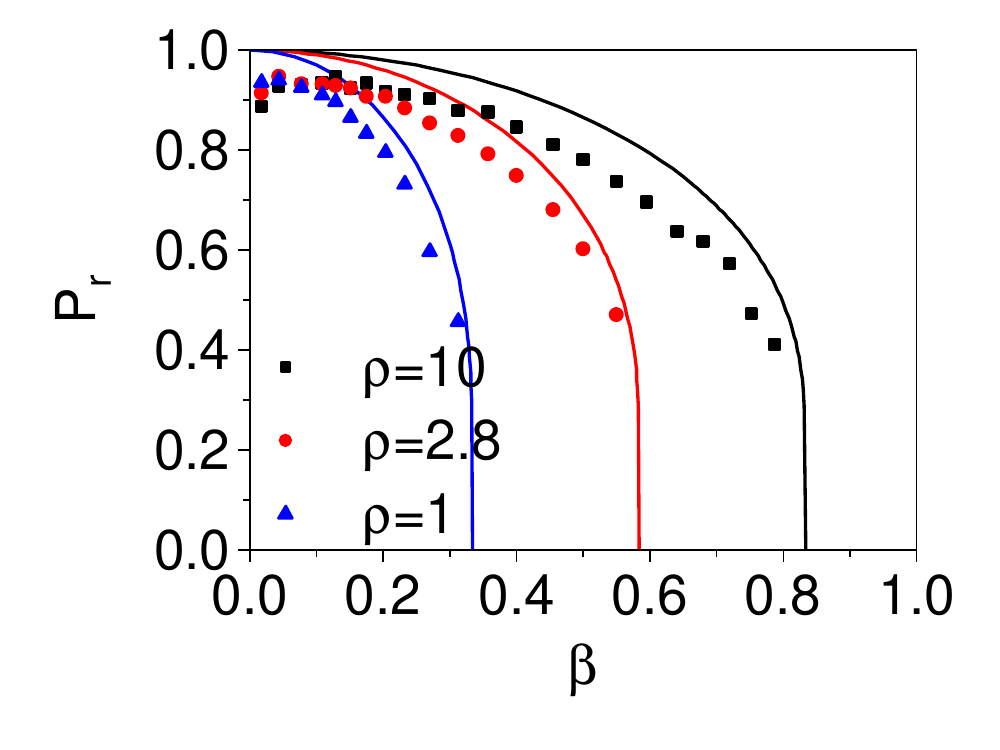}
	\caption{Power ratio reference to $P_0=V_{\rm{o}}^2/R_{\rm{L}}$.}\label{fig3}
\end{figure}

In fact, the smaller measured $P_{\rm{r}}$ is due to the voltage loss of the selected Op-Amp: when the Op-Amp is powered by DC voltage $V_{\rm{CC}}$, the actual output voltage of the Op-Amp is 15\% less than the maximal possible output voltage $V_{\rm{CC}}$. Alternatively, if we calculate the power ratio reference to $P_0=V_{\rm{o}}^2/R_{\rm{L}}$ where $V_{\rm{o}}$ is the actual output voltage of the Op-Amp, the updated results are shown as the solid symbols in Fig.~\ref{fig3}. As we observe, the measured power ratio with new reference is much closer to the analytical formula Eq.~\ref{eqPr}.

We note that, in Fig.~\ref{fig2} and \ref{fig3}, the boundaries at larger $\beta$ values are given by the self-oscillating condition Eq.~\ref{eqcond}. On the other hand, from the power ratio figures it is clear that smaller $\beta$ values can give higher power ratios, i.e., more power delivered to the load. Moreover, from Eq.~\ref{eqcond}, we know that smaller $\beta$ supports more $\rho$ values which means wider $C$ ranges. Therefore, smaller $\beta$ is preferable for real implementations. 

\subsection{Varying Load Resistance $R_{\rm{L}}$}
\begin{figure}[t!]
	\centering
	\subfigure[]
	{\includegraphics[width=0.24\textwidth]{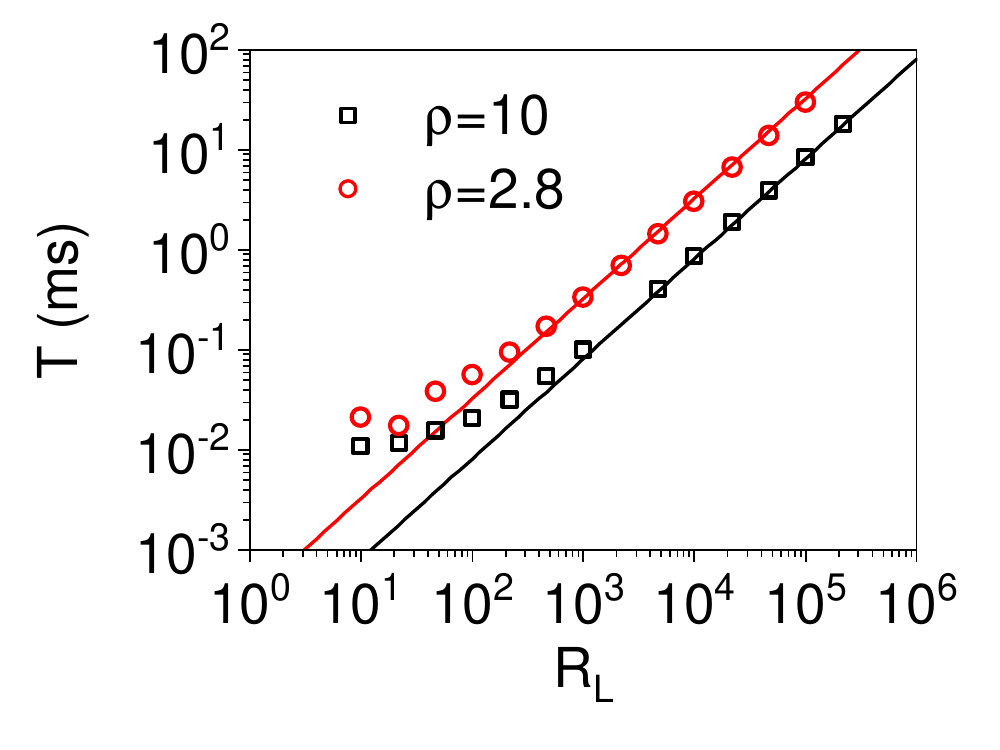}\label{fig4a}}
	\centering
	\subfigure[]
	{\includegraphics[width=0.24\textwidth]{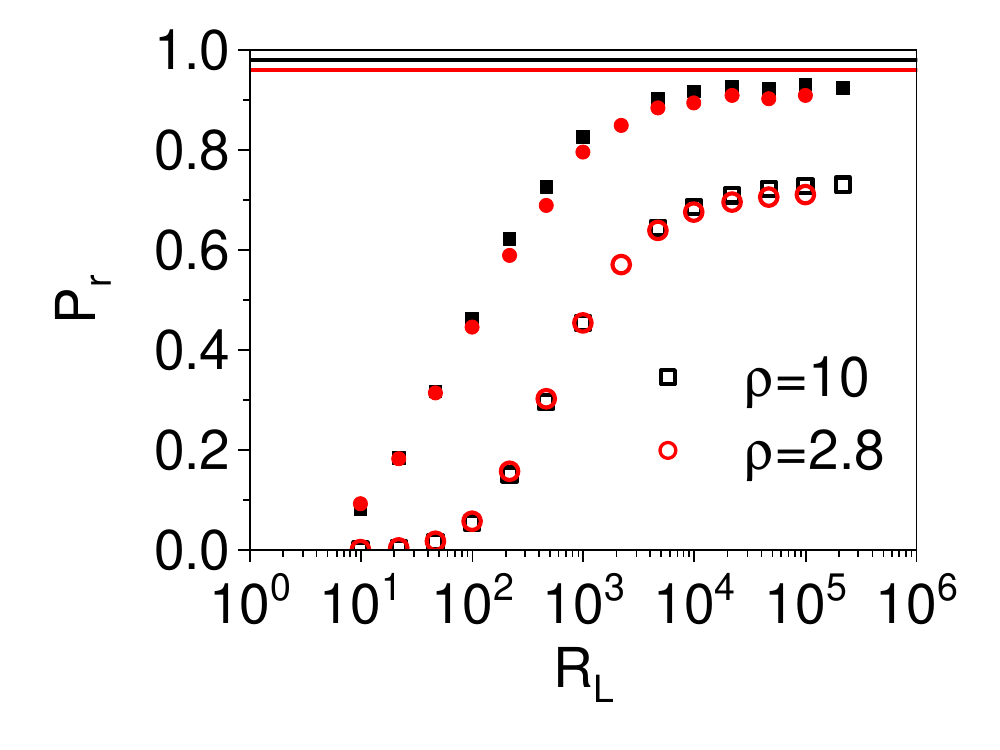}\label{fig4b}}
	\caption{(a) Oscillation period $T$ and (b) power ratio $P_{\rm{r}}$ on the load (the open and solid symbols have the same definition as those in Fig.~\ref{fig2b} and \ref{fig3}) when varying $R_{\rm{L}}$ for different $\rho$ and $C_1$ configurations.}
	\label{fig4}
\end{figure}

Next, we study the performance of the proposed WPT design under the change of the load resistance $R_{\rm{L}}$. In this case, we fix $\beta=0.2$ and choose two configurations: $\rho=10$ with $C_1=100$~nF and $\rho=2.8$ with $C_1=356$~nF. The experimentally measured period $T$ and the power ratio $P_{\rm{r}}$ are shown as the open symbols in Fig.~\ref{fig4}. It is clear that the measured $T$ follows the analytical prediction for large $R_{\rm{L}}$ while reaches a limit around $10^{-2}$ ms when $R_{\rm{L}}$ gets smaller. It is actually due to the finite slew rate of the chosen Op-Amp, meaning that the Op-Amp cannot follow the fast change of the output. For the same reason, the power ratio $P_{\rm{r}}$ drops down with smaller $R_{\rm{L}}$, while in theory $P_{\rm{r}}$ has the constant values $0.98$ and $0.96$ for the two configurations (black and red lines in Fig.~\ref{fig4b}) as $P_{\rm{r}}$ is not a function of $R_{\rm{L}}$.

\section{Conclusion}
Based on the proposed self-oscillating circuit, we have shown that the self-oscillating capacitive WPT design \cite{Radi2017arxiv} is robust under the change of the feedback circuit, especially for the load resistance $R_{\rm{L}}$ and capacitance $C$ which is embedded in the parameter $\rho$. These results are very different from the conventional designs which can only operate effectively in a narrow frequency range and dynamic adjustment is required for maintaining matching when the working conditions change. The next steps of this work is to study the robustness against the change of $\rho$ ($C$ embedded, modeling changing transfer distance) and replace $C$ with real metal plates.

\section*{ACKNOWLEDGEMENT}
This work was supported by the European Union’s Horizon 2020 research and innovation programme-Future Emerging Topics (FETOPEN) under grant agreement No 736876.


\begin{thebibliography}{1}

\bibitem{Dai2015survey}
J. Dai, and D. C. Ludois, ``A survey of wireless power transfer and a critical comparison of inductive and capacitive coupling for small gap applications'', \emph{IEEE Trans. Power Electron.}, vol. 30, pp. 6017--6029, November 2015.

\bibitem{Fan2017PT}
S. Assawaworrarit, X. Yu, and S. Fan, ``Robust wireless power transfer using a nonlinear parity-time-symmetric circuit'', \emph{Nature}, vol. 546, pp. 387--390, June 2017.

\bibitem{Radi2017arxiv}
Y. Ra'di, B. Chowkwale, C. Valagiannopoulos, F. Liu, A. Al\`{u}, C. R. Simovski, and S. A. Tretyakov, ``On-site wireless power generation'', \emph{IEEE Trans. Antennas Propag.}, vol. 66, pp. 4260--4268, August 2018.
\end{thebibliography}
\end{document}